\documentclass[journal=jpclcd,manuscript=letter,layout=twocolumn]{achemso}
\usepackage{multicol}
\usepackage{url}
\usepackage{hyperref}
\usepackage{caption}
\usepackage{amsmath}
\usepackage[dvipsnames]{xcolor}
\usepackage{graphicx}

\title[An \textsf{achemso} demo]
  {Efficient Cysteine Conformer Search with Bayesian Optimization}

\usepackage{xcolor}
\usepackage{graphics,graphicx}
\definecolor{aaltoOrange}{RGB}{255,121,0}%

\author{Lincan Fang}

\author{Esko Makkonen}

\author{Milica Todorovi\'{c}}

\author{Patrick Rinke}
\email{patrick.rinke@aalto.fi}

\author{Xi Chen}
\email{xi.6.chen@aalto.fi}

\affiliation{Department of Applied Physics, Aalto University, 00076 AALTO, Finland}

\begin{document}
\maketitle

%%%%%%%%%%%%%%%%%%%%%%%%%%%%%%%%%%%%%%%%%%%%%%%%%%%%%%%%%%%%%%%%%%%%%
%% The document title should be given as usual. Some journals require
%% a running title from the author: this should be supplied as an
%% optional argument to \title.
%%%%%%%%%%%%%%%%%%%%%%%%%%%%%%%%%%%%%%%%%%%%%%%%%%%%%%%%%%%%%%%%%%%%%

%%%%%%%%%%%%%%%%%%%%%%%%%%%%%%%%%%%%%%%%%%%%%%%%%%%%%%%%%%%%%%%%%%%%%
%% Some journals require a list of abbreviations or keywords to be
%% supplied. These should be set up here, and will be printed after
%% the title and author information, if needed.
%%%%%%%%%%%%%%%%%%%%%%%%%%%%%%%%%%%%%%%%%%%%%%%%%%%%%%%%%%%%%%%%%%%%%
%\abbreviations{IR,NMR,UV}
%\keywords{American Chemical Society, \LaTeX}

%%%%%%%%%%%%%%%%%%%%%%%%%%%%%%%%%%%%%%%%%%%%%%%%%%%%%%%%%%%%%%%%%%%%%
%% The manuscript does not need to include \maketitle, which is
%% executed automatically.
%%%%%%%%%%%%%%%%%%%%%%%%%%%%%%%%%%%%%%%%%%%%%%%%%%%%%%%%%%%%%%%%%%%%%

%%%%%%%%%%%%%%%%%%%%%%%%%%%%%%%%%%%%%%%%%%%%%%%%%%%%%%%%%%%%%%%%%%%%%
%% The "tocentry" environment can be used to create an entry for the
%% graphical table of contents. It is given here as some journals
%% require that it is printed as part of the abstract page. It will
%% be automatically moved as appropriate.
%%%%%%%%%%%%%%%%%%%%%%%%%%%%%%%%%%%%%%%%%%%%%%%%%%%%%%%%%%%%%%%%%%%%%
\makeatletter
\setlength\acs@tocentry@height{9.0cm}
\setlength\acs@tocentry@width{3.5cm}
\makeatother
\begin{tocentry}

\begin{center}
\includegraphics[width=9.0cm]{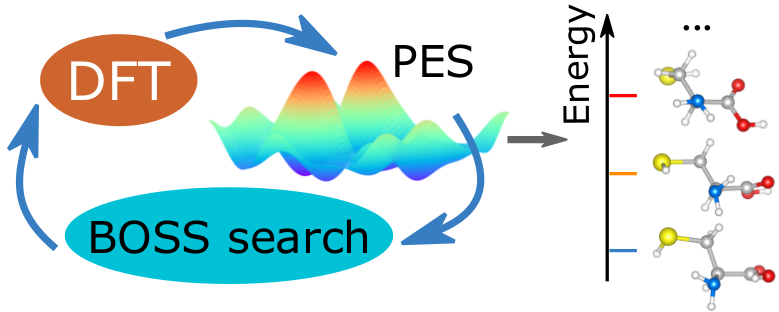}
\end{center}

\end{tocentry}

\twocolumn[
\begin{@twocolumnfalse}
\begin{abstract}

Finding low-energy molecular conformers is challenging due to the high dimensionality of the search space and the computational cost of accurate quantum chemical methods for determining conformer structures and energies. Here, we combine active-learning Bayesian optimization (BO) algorithms with quantum chemistry methods to address this challenge. Using cysteine as an example, we show that our procedure is both efficient and accurate. After only one thousand single-point calculations and approximately thirty structure relaxations, which is less than 10\% computational cost of the current fastest method, we have found the low-energy conformers in good agreement with experimental measurements and reference calculations.
\end{abstract}
\end{@twocolumnfalse}
]

%Identifying low-energy molecular conformers is challenging due to the high dimensionality of the search space and the computational cost of accurate quantum chemical methods for determining conformer structures and energies. Here, we combine active-learning Bayesian optimization (BO) algorithms with quantum chemistry methods to address this challenge. Using cysteine as an example, we demonstrate how 5-dimensional potential energy landscapes can be constructed with modest sampling effort, and the local minima extracted to refine numerous conformer structures. This approach accurately reproduced all experimentally measured conformer structures, with energy rankings in excellent agreement with MP4 reference calculations.

\section{Introduction}

%%%%%%%%%

A molecular conformer is a distinct conformation corresponding to a minimum on the molecule's potential energy surface (PES). Any molecule with rotatable bonds has several stable conformer structures, each associated with different chemical and electrical properties. At ambient temperatures, all the properties of that molecule are the combination of the properties of its conformers accessible at the temperature of the study\cite{confer1}. Therefore, identifying the low-energy conformers and determining their energy ranking continues to be a topic of great interest in computational chemistry\cite{confer-CC}, cheminformatics \cite{confer2,confer3}, computational drug design\cite{confer-drug}, and structure-based virtual screening\cite{confer-vs}. While one configuration of a small molecule can be simulated routinely by \textit{ab initio} methods, the large size of configurational phase space and the considerable number of local minima in typical energy landscapes make conformer searches one of the persistent challenges in molecular modeling\cite{confer1,confer2,confer3}.

The first challenge in conformer search is sufficient sampling of the configurational space. The conformational space (bond lengths, bond angles, and torsions) for even relatively small molecules is enormous. Certain approximations are necessary to make the problem tractable. Since the bond lengths and angles are relatively rigid in molecules, and the different conformers originate from the flexible rotational groups, most search methods focus on sampling the torsion angles in molecules, while keeping bond length and angles fixed \cite{confer1}. A variety of methods and tools have been developed to generate diverse
conformer structures\cite{T1LC,T2LC,T3LC,T4LC,T6LC,T7LC}. These methods can be broadly classified to be either systematic or stochastic. 

A systematic method relies on a grid to sample all the possible torsion angles in the molecule. This approach is deterministic, but limited to small molecules because it scales poorly with increasing numbers of relevant torsion angles, i.e. search dimensions.
Stochastic methods randomly sample the torsion angle phase space (sometimes restricted to predefined, most relevant ranges) based on different algorithms such as Monte Carlo annealing \cite{MCAL1LC,MCAL2LC}, minima hopping \cite{2004minima}, basin hopping \cite{2015basin}, distance geometry \cite{distance-geoLC} and genetic algorithms \cite{GENIC1LC,T1LC}. Stochastic methods can be applied to larger molecules with high-dimensional conformer space, but the predicted conformers may vary. Extensive sampling is required and the results may depend on the random nature of the process.

Knowledge-based methods have also been developed \cite{KNBASED2LC,KNBASED1LC} to achieve more consistent results. They use a predefined library for torsion angles and ring conformations. The library is typically based on experimental structures in databases such as the Cambridge Structure Database (CSD) \cite{CSDLC} or the Protein Data Bank (PDB) \cite{PDBLC}. To search the conformers, knowledge-based methods usually need to be combined with the different systematic or stochastic algorithms mentioned before.

The second challenge in conformer searches is the sufficiently accurate mapping of energies and structures. Two classes of total energy approaches are commonly used:
force field-based methods and quantum chemistry methods such as density functional theory (DFT) and coupled cluster (CC).  Quantum chemistry methods achieve higher accuracy in the estimation of molecular energies than force fields because they describe the interactions and polarization in molecules more accurately. However, they are computationally costly. More often than not, quantum chemistry methods are too expensive to provide energies for all configurations produced in the search. 

To balance efficiency and accuracy, hierarchical methods have been developed. People employ  fast computational methods with lower accuracy to scan configurational space, then funnel promising candidate structures through more costly methods with  higher accuracy to refine the conformer structures and energies (such as force fields $\rightarrow$ DFT \cite{hiera1,hiera2} or HF $\rightarrow$ MP2 $\rightarrow$ CCSD(T)\cite{Theo3-Cys}).  Methods at different levels predict different PES of molecules. To avoid missing the true low-energy conformers, a large portion of configurational space have to be sampled at a lower accuracy method level, and many structures need to be optimized at a higher level.

In recent years, artificial intelligence (AI) and machine learning (ML) techniques such as genetic algorithms\cite{ML-M1,ML-M2}, artificial neural network\cite{ML-M3,ML-NN}, and machine learned force fields\cite{ML-FF} were used to accelerate structure-to-energy predictions for molecules.  The majority of these schemes requires a large number of data points, which may be costly to compute with \textit{ab-initio} methods. To reduce the amount of required data, Bayesian optimization was introduced in the structure search\cite{BO,BOSS}.  Bayesian optimization search schemes belong to the active learning family of methods, which generate data on the fly for optimal knowledge gain. 

\begin{figure}
\centering
\includegraphics[width= 6.5cm]{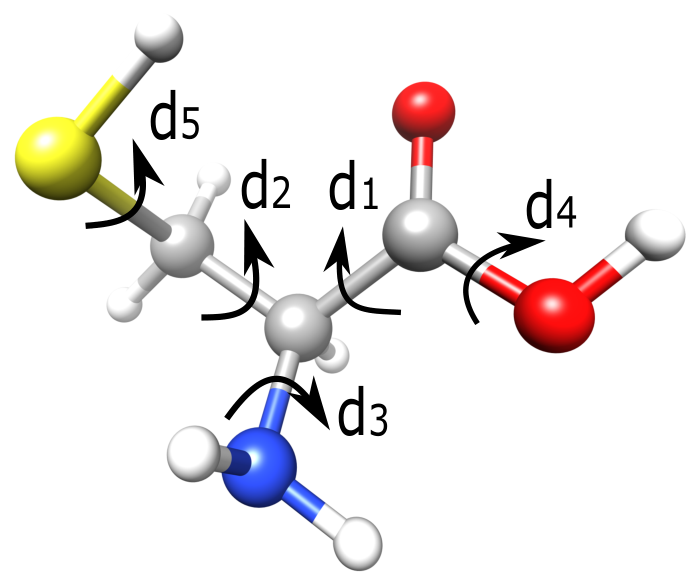}
\caption{Ball-and-stick model of the cysteine molecule. Red is used for oxygen, white for hydrogen, gray for carbon, blue for nitrogen, and yellow for sulfur. $d_{1}, d_{2}, d_{3},d_{4},d_{5}$ label the five  dihedral angles of cysteine that we use to define our search space.}
\label{fig:structure}
\end{figure}

In this article, we present a new working procedure for molecular conformer identification and ranking. We combined the Bayesian Optimization Structure Search (BOSS) approach \cite{BOSS} and quantum chemistry simulations to find the conformers of small molecules and accurately predict their relative stability. BOSS is a python-based tool for global phase space exploration based on Bayesian optimisation \cite{BOSS-web}. Beyond the Bayesian active learning method for the global minimum conformer search in the Ref.~\citenum{BO}, our procedure aims to find all the relevant conformers in one run. We use cysteine as a model system to demonstrate our methodology.

Cysteine was chosen as the model system for several reasons. First, it is an amino acid with critical biological functions. Second, it is the only amino acid that has a -SH group. The strong S-Ag and S-Au bonds make it uniquely interesting for use in hybrid nanomaterials \cite{Au-S,Au-S2}.  Third, cysteine has five rotational groups, as shown in Figure \ref{fig:structure}. Therefore it is an interesting and accessible 5-dimesional (5-D) system for Bayesian optimization. Last, the structures and the energy order of cysteine's conformers have been calculated by several groups using the grid sample method \cite{Theo1-Cys, Theo2-Cys,  Theo3-Cys} and characterized by Fourier transform microwave spectroscopy experiments\cite{Exp} so that we can compare the accuracy and efficiency of our new procedure with other computational and experimental results.  

In brief, using cysteine as an example, we present an efficient and reliable procedure to predict the structures and energies of molecular conformers. BOSS ensures sufficient sampling of the configurational phase space and outputs the structures associated with local energy minima. We post-process the machine-learned conformer candidates with geometry optimization and then add free energy corrections to obtain the final ranking. We tested the effect of different exchange-correlation functionals and van de Waals interactions on the ranking order. Finally, we applied coupled cluster corrections to the lowest-energy conformers. The methods and results will be presented in the following sections.

\section{Methods}

\subsection{A. BOSS-based Molecular Conformer Search}

\begin{figure*}[hbt!]
\includegraphics[width=16 cm]{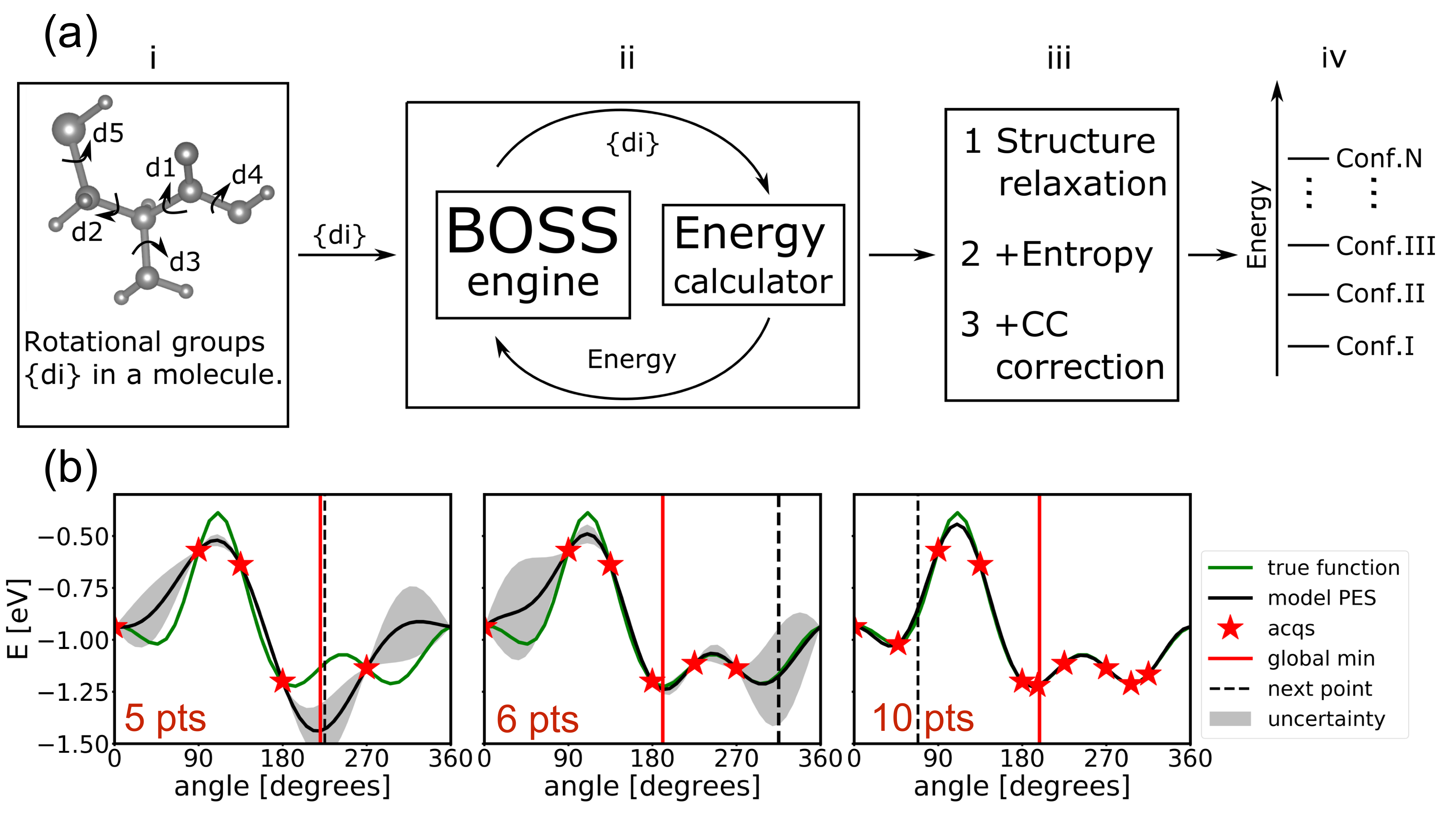}
\centering
\caption{\label{fig:overview}{(a) Overview of our BOSS-based procedure for molecular conformer search, featuring \textbf{i} System Preparation, \textbf{ii} Bayesian Optimization Conformer Search, \textbf{iii} Refinement and \textbf{iv} Validation. (b) BOSS iterative inference of a one-dimensional (1D) PES of the d$_1$ dihedral  angle  of  cysteine. The GP's native uncertainty (gray areas) facilitates exploratory data sampling. The global minimum location and the entire PES are learned in 10 data acquisitions. }}
\end{figure*}

Our BOSS-based procedure for molecular conformer search contains four steps: \textbf{i} System Preparation, \textbf{ii} Bayesian Optimization Conformer Search, \textbf{iii} Refinement and \textbf{iv} Validation, as illustrated in Figure \ref{fig:overview}a.

In step \textbf{i}, we first obtain an xyz-file of our molecule of interest from a database and then perform a single geometry optimization with a quantum chemistry method. Then we calculate the z-matrix to find the dihedral angles. We chose the dihedral angles d$_n$ to describe the different conformers, as they are typically the most informative degrees of freedom for conformers. We keep all bond lengths and angles fixed at their optimized values. Such dimensionality reduction is standard practice to expedite the molecular conformer search, as mentioned in the introduction. 

In step \textbf{ii}, BOSS actively learns the PES of the molecule by Bayesian optimization iterative data sampling. Each data ``point'' consists of the set of dihedral angles {d$_n$} for a molecular configuration and its corresponding total energy $E$. In this step, we use DFT as the calculator. $E$ therefore corresponds to the DFT total energy of a molecular configuration.

BOSS employs Gaussian Process (GP) models \cite{GP-book} to fit a surrogate PES to the data points, then refines it by acquiring more data points at locations which minimize the exploratory Lower Confidence Bound (eLCB) acquisition function\cite{BOSS-web}.  The most-likely PES model for given data is the GP posterior mean. The lack of confidence in the model is reflected by the GP posterior variance, which vanishes at the data points, and rises in unexplored areas of phase space. 
The key concepts of this active learning approach are illustrated in Figure \ref{fig:overview}b, in which BOSS iteratively infers a one-dimensional PES of the d$_1$ dihedral angle of cysteine. The global minimum location and the entire PES are learned in 10 data acquisitions. In analogy with the 1D example, BOSS actively learns the PES in N dimensions until convergence is achieved. The advantage of BOSS is not only its efficiency, but also the fact that it explores both the global minimum and local minima of the PES during the search. We exploit this feature to find conformers beyond the global minimum, which we associate with the local minima of the PES. A more detailed introduction of the Bayesian optimization method can be found in Refs.~\citenum{BOSS} and \citenum{BOSS-new}.

After the BOSS-predicted PES has converged, in step \textbf{iii}, we analyze the PES to extract the local minima locations and related structures. Since the PES and its gradients can be computed efficiently at any location in the N-dimensional phase space from the GP model, BOSS post-processing routines perform multiple L-BFGS (limited-memory Broyden-Fletcher-Goldfarb-Shanno algorithm) minimizations, using the locations of the data acquisitions as starting points. The resulting list of minima is processed to remove duplicates. Using cysteine as an example, if local minima structures have the same types of -CH$_2$SH chain and hydrogen bonds, we consider them as duplicates and the higher energy structures will be purged. 

The conformer set is next refined by geometry optimization and entropy corrections. First, all degrees of freedom (including bond lengths and angles) are relaxed to obtain optimized structures and energies. Next, we add vibrational entropy corrections following previous studies\cite{vib2,vib3}. We compute and add the zero-point energy and the vibrational free energy at 300K to the energies of optimized conformers. 

In step \textbf{iii} we also go beyond DFT. We perform a coupled cluster calculation for the DFT-optimized conformer structures in a relevant energy window. The difference between the coupled cluster and DFT total energy, here called CC correction, is then added to the entropy corrections we added earlier in step \textbf{iii}.

In step \textbf{iv}, we validate our results by comparing the low-energy conformers we found  to  experimental and other computational results. System preparation and final validation require human input, but procedures featuring structure search and refinement can be made fully automated into a computational workflow. 

\subsection{B. Computational Methods}

In this work, we employed DFT as the predominant energy calculator and employed the all-electron codes FHI-aims\cite{FHI-AIMS,Aims-efficientLC,AIMS-HYBcl} for all DFT calculations. "Tight" numerical settings and "tier 2" basis sets were used throughout.  To investigate the influence of the exchange-correlation functional and the level of dispersion correction on the final results, we performed our conformer search with the PBE+TS\cite{PBE,TS}, PBE+MBD\cite{PBE,MBD}, PBE0+TS \cite{PBE0,TS} and the PBE0+MBD \cite{PBE0,MBD} functionals. For geometry optimizations, the geometry was considered to be converged when the maximum residual force was below 0.001 $\mathrm{eV/\AA}$.  

Vibrational free energies were computed using the finite-difference method within the harmonic approximation. We used a finite-difference displacement length of $\mathrm{\delta = 0.0025 \AA}$. The vibrational free energy $F_{vib}$ was then calculated as follows
\begin{equation}
\begin{split}
    F_{vib} (T) & =\int d\omega g(\omega)\frac{\hbar\omega}{2} \\
    & +\int d\omega g(\omega)k_BTln[1-exp(-\frac{\hbar\omega}{k_BT})]
\end{split}
\end{equation}
where $g(\omega)$ is the phonon density of states and T, $\omega$ and $k_B$ are temperature, frequency and Boltzmann constant.

Going beyond DFT, we performed CC calculations with single, double and perturbative triple excitations (CCSD(T)). These were done as single-point calculations using the structures from the PBE0+MBD calculation with aug-cc-pVTZ basis sets. For validation purposes, we also performed MP4 single-point calculations for selected conformers in their PBE0+MBD geometries with 6-311++G(d,p) and aug-cc-pVTZ basis sets. We used the  Gaussian16 code \cite{g16} for the CCSD(T) and MP4 simulations.

To support open data-driven chemistry and materials science \cite{Himanen/Geurts/Foster/Rinke:2019}, we uploaded all calculations of this work to the Novel Materials Discovery (NOMAD) laboratory \cite{NOMAD}.

\begin{figure}[hbt!]
\centering
\includegraphics[width=7.5cm]{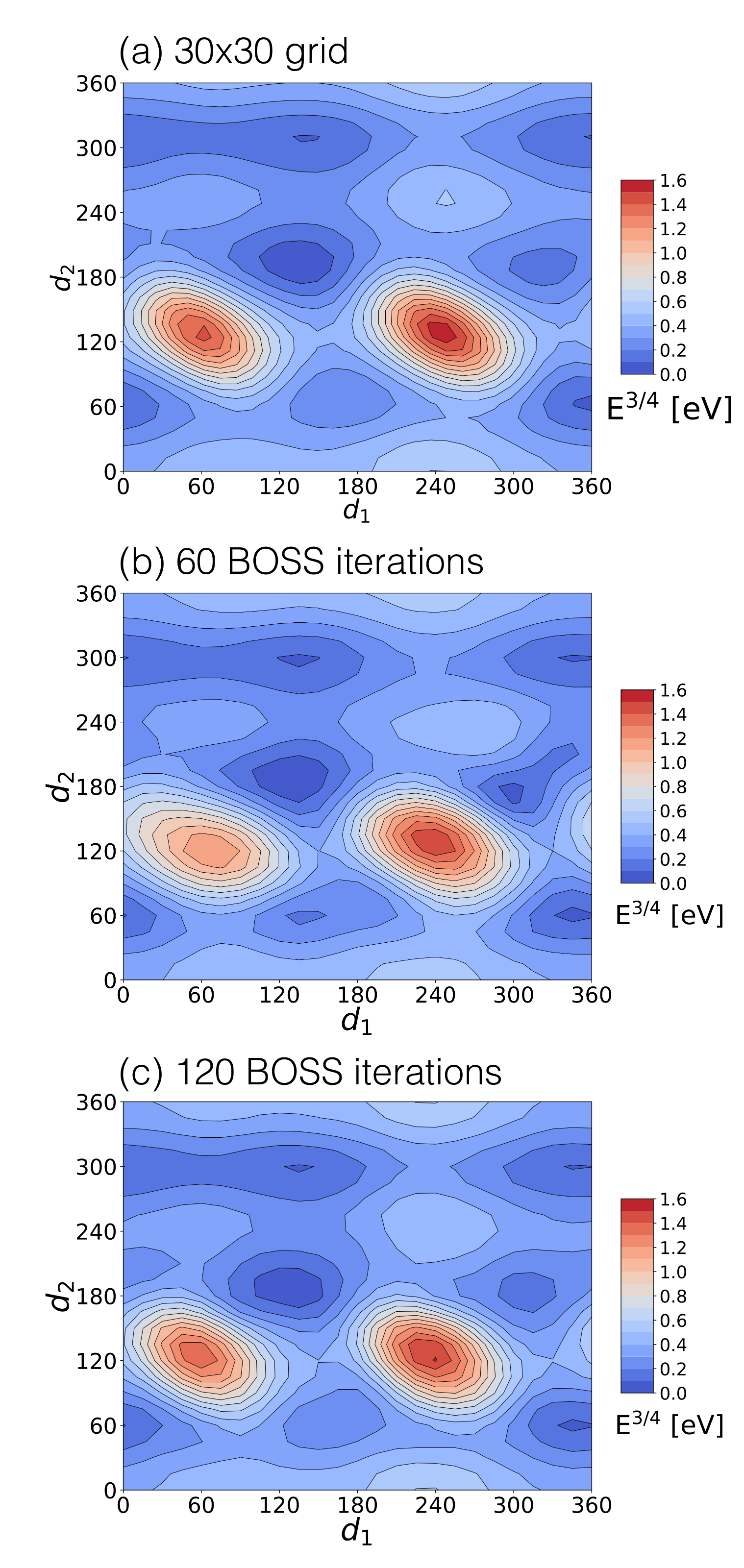}  
\caption{The 2-D ($d_1$, $d_2$)  PES map of cysteine generated by (a) 30$\times$30=900 DFT single-points energy calculations, (b) 60 BOSS iterations, and (c) 120 BOSS iterations. To increase the PES contrast, E$^{3/4}$ instead of E is plotted. }
\label{fig:2d}
\end{figure}

\subsection{C. 2-D test}
To test the accuracy and efficiency of step \textbf{ii} in our procedure, we started with a 2-D search case in cysteine (Figure \ref{fig:2d}). First we rotated the $d_1$ and $d_2$ dihedral angles to generate a reference map on a fine grid (30 $\times$ 30 points, Figure \ref{fig:2d}a). Then $d_1$ and $d_2$ were  sampled by BOSS. In both approaches, the bond lengths, bond angles and other dihedral angles ($d_3$ =180.03, $d_4$ = 145.59 , $d_5$=180.03) were fixed in their DFT-optimized values. We obtained the energy of each structure with  single-point PBE0+MBD calculations and then plot the energy relative to the global minimum. 

The 2-D PES maps after 60 and 120 data acquisitions are shown in Figure \ref{fig:2d}b and \ref{fig:2d}c. Looking at Figure \ref{fig:2d}, we find that BOSS captures  correct minima and maxima already after 60 data acquisitions (6\% of the computational cost of the grid method), while after 120 data acquisitions the BOSS PES resembles the reference map very well. This 2D PES features 6 energy minima of similar depth, suggesting considerable complexity of cysteine conformational phase space and many competing minima. We apply abundant sampling in high-dimensional problems so that we can recover all relevant conformer solutions.

\subsection{D. Cysteine Conformer Search in 5-D }
After demonstrating the BOSS rationale in 1-D and 2-D, we proceed to 5 dimensions. The five dihedral angles ($d_1$ - $d_5$) in cysteine were sampled simultaneously by BOSS and the energies of the corresponding configurations were evaluated with the PBE0+MBD functional. 

\begin{figure}[hbt!]
\centering
\includegraphics[width= 8 cm]{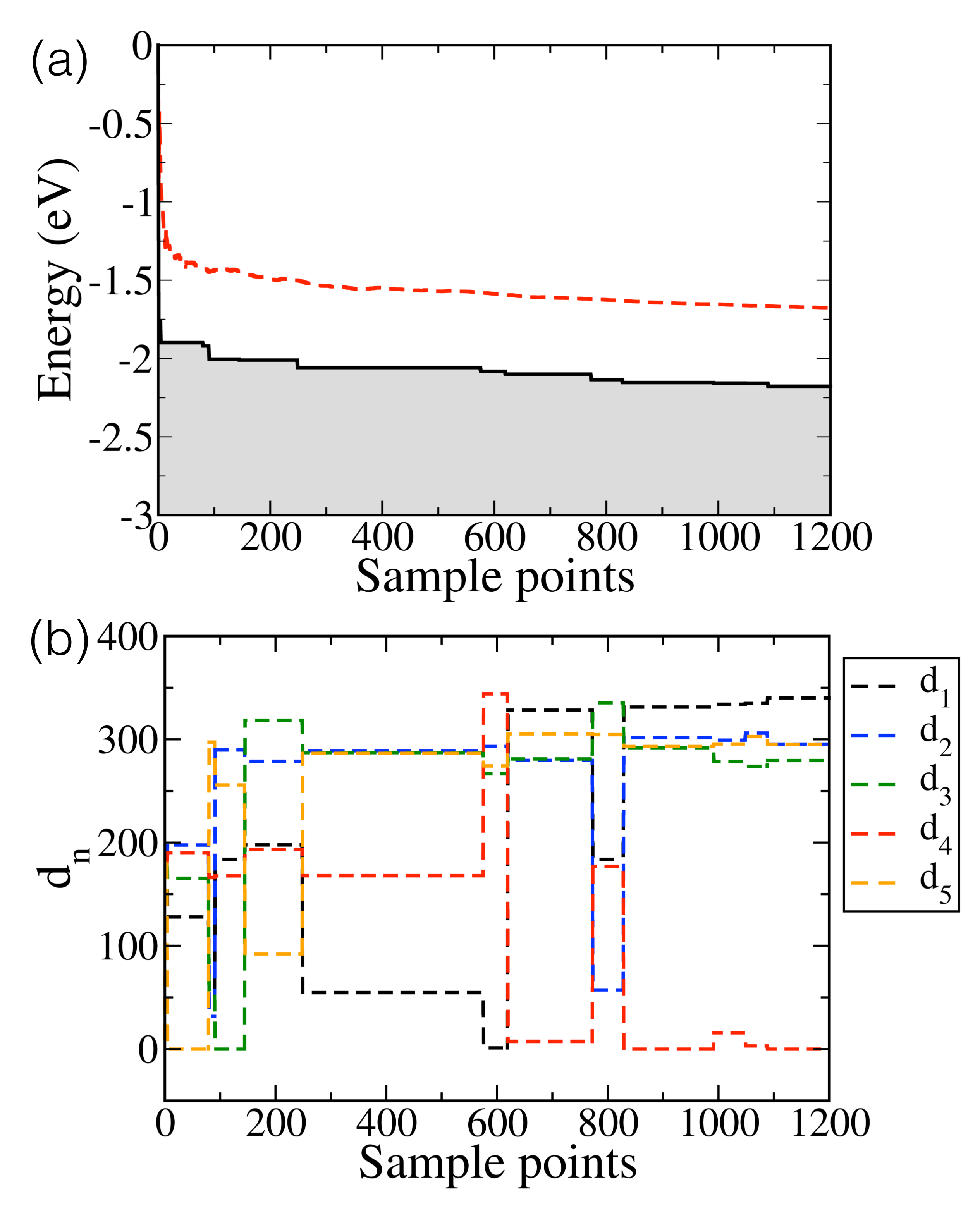}
\caption{(a) Convergence of the global minimum energy computed from the BOSS predicted global minimum configuration (black line). The average computed energy of the sampled conformers is shown with red dashed line. (b) Value of the dihedral angles d$_n$ of the BOSS predicted global minimum as a function of the number of sampled points.}
\label{fig:err}
\end{figure}

Figure \ref{fig:err} illustrates the refinement of the predicted global minimum with iterative configurational sampling. The lowest observed energy (calculated from the BOSS-predicted global minimum conformer) is shown in Figure \ref{fig:err}a and the values of the corresponding dihedral angles d$_n$  in Figure \ref{fig:err}b. The lowest energy observed decreases continuously. Throughout the procedure, the geometry of the 
global minimum 
conformer changes, as Figure \ref{fig:err}b illustrates. The global minimum undergoes several refinements until, at iteration 830, both the energy and the dihedral angles are converged and  only have negligible changes ($\Delta E<$ 0.025 eV and $\Delta d<$ 10$^\circ$). 

Improvements of the global minimum prediction is due to instances of visiting low energy configurations chosen smartly form a vast 5-D space. However, most model refinements proceeded with higher energy conformers and explores local minima of the PES, on average in the region 0.4 eV above the predicted global minimum, as shown by the red dashed line in Figure \ref{fig:err}a.

Next we address the convergence of the low energy part of the PES. This is not a trivial task, as we cannot monitor the PES in every point of 5-D space. It also turns out to be impractical to track the dihedral angles of several low energy conformers and monitor convergence as we did for the global minimum. The reason is that many conformers are very close in energy and switch order as the iterations progress. We therefore decided to take the energy-versus-conformer-number curve as convergence indicator.

\begin{figure}[hbt!]
\centering
\includegraphics[width= 8 cm]{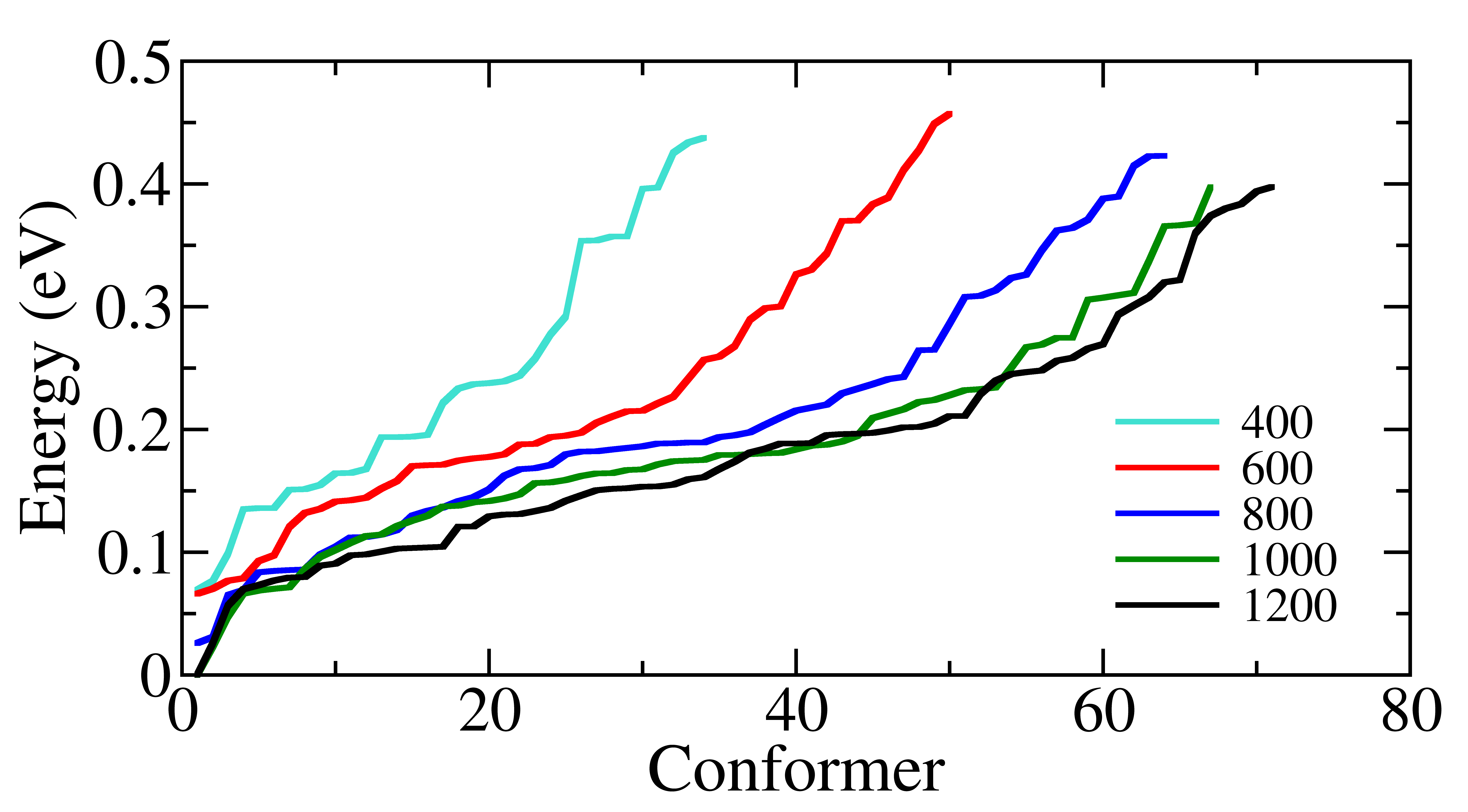}
\caption{Progression of the relative energy of predicted local-minima for a PBE0+MBD BOSS run with a total number of 1200 iterations. Shown are intermediate curves at 400, 600, 800 and 1000 iterations.}
\label{fig:BOSS-energy}
\end{figure}

Figure~\ref{fig:BOSS-energy} shows the relative energy to the global minimum for all minima found up to 0.4 eV after 400, 600, 800, 1000 and 1200 BOSS iterations. In the figure, 0 eV is set to be the lowest energy found in the 1000-th iteration. The curves after only 400 and 600 iterations still rise steeply and feature the wrong global minimum (i.e. do not start at 0 eV). With increasing number of iterations, the curves flatten and gradually approach the curve for 1200 iterations. At 1000 iterations the curve is very similar to that of 1200 iterations, which suggests that not only the global but also the local minima are converged. The 2-D (d1, d2) projected PES maps are presented in Figure S1.

To reliably identify even higher-energy PES minima, BOSS post-processing routines conduct an exhaustive minimiser-based search. Consequently, the resulting local minima list may contain similar structures or duplicates. In this study, if two local minima structures have the same types of -CH$_2$SH chain ($\Delta d_2$$<$10$^{\circ}$) and hydrogen bonds 
($\Delta d_4$$<$10$^{\circ}$, $\Delta d_1$$<$30$^{\circ}$, $\Delta d_3$$<$30$^{\circ}$), we consider them as duplicates and only keep the lowest energy one for further processing.

Next we optimize the structures and include the vibration energy as described in Section 2A. Finally, we apply CCSD(T) single-point corrections to the 15 lowest energy conformers obtained from the PBE0+MBD calculations.

\section{Results and Discussion}
Using the methodology introduced in the previous sections, we performed four independent conformer searches with the PBE+TS, PBE+MBD, PBE0+TS and the PBE0+MBD functionals for cysteine. In this section, we systemically assess how the different exchange-correlation functionals and van de Waals corrections affect the results and discuss how the different steps improve accuracy. We also compare our predictions with the experimental results and reference calculations \cite{Exp,Theo3-Cys}.

We choose two references to make the comparison and validate our results. Ref~1\cite{Exp} contains both experimental and computational results. The computational energy ordering is obtained from single-point MP4 calculation on MP2 optimized structures using 6-311++G(d,p) basis sets. In the reference, six experimental conformers were found by rotational spectroscopy (labeled in red in Figure \ref{fig:conformers}); five other low-energy conformers were predicted from the MP4 simulations, but were not detected in the experiment (labeled in black in Figure \ref{fig:conformers}). The authors of Ref~2\cite{Theo3-Cys} did a systematic scan of 11644 initial structures at the HF/3-21G level,
located 71 unique conformers of cysteine using the MP2(FC)/cc-pVTZ method and finally determined the relative energy of the eleven lowest-energy conformers with CCSD(T). Ref 2 also provides xyz-coordinates for the observed conformers.

\subsection{Conformer hierarchy}

\begin{figure*}[hbt!]
\includegraphics[width=15 cm]{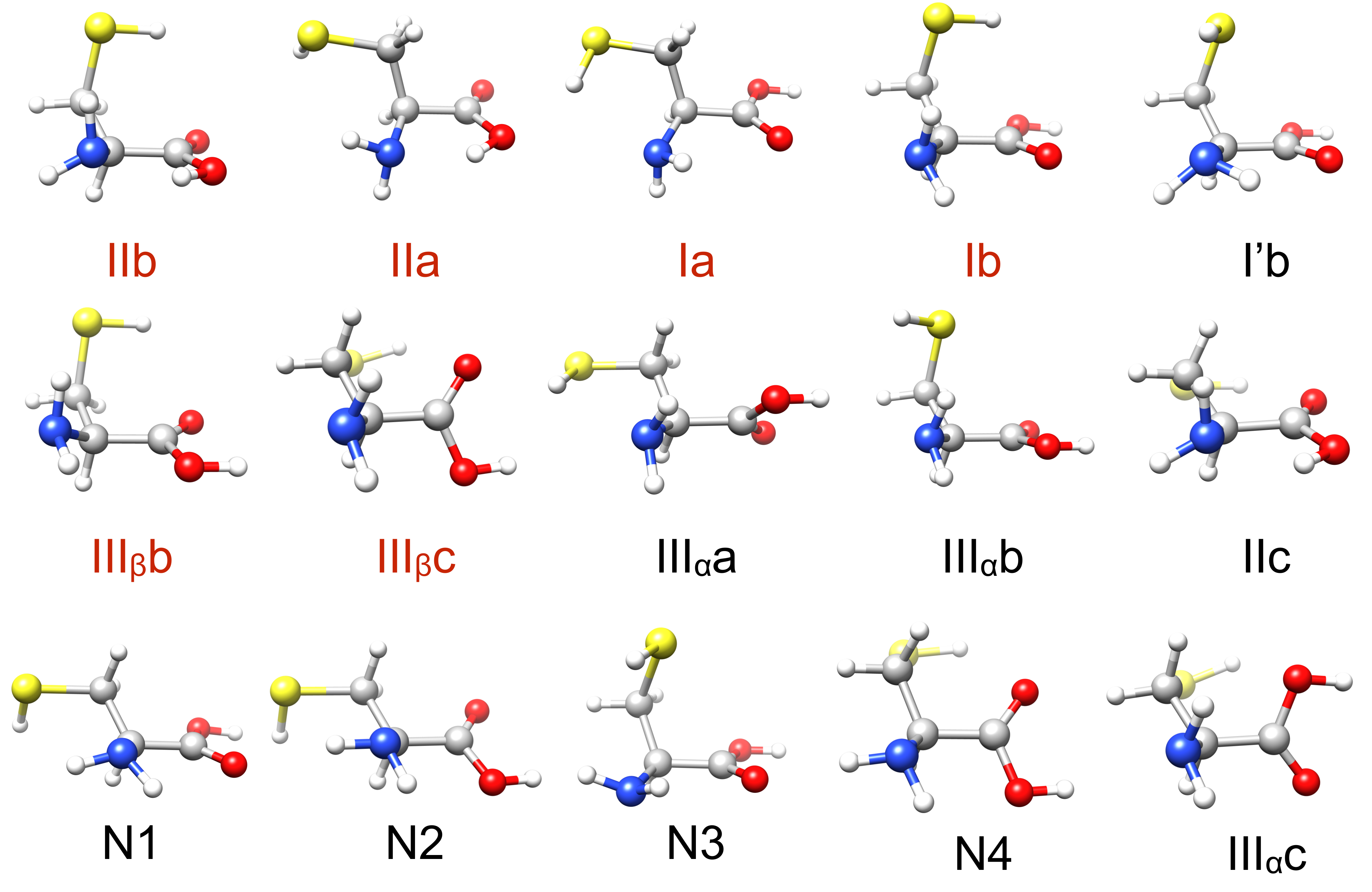}
\centering
\caption{\label{fig:conformers} Predicted low energy conformers of cysteine from the PBE0+MBD search. Conformers are named as I (NH--O=C), II (OH--N), and III (NH--OH) depending on the type of the hydrogen bonds, and as a, b, or c depending on the configuration of the $\mathrm{-CH_{2}SH}$ side chain, following Ref~1\cite{Exp}. The experimentally detected conformers are marked in red and other conformers marked in black. The colour scheme of the atoms is the same as in Fig.~\ref{fig:structure}.}
\end{figure*}

The predicted 15 lowest energy conformer structures of cysteine with the PBE0+MBD functional are shown in Figure \ref{fig:conformers}. The atomic coordinates of the conformers can been found in the Supplementary Material. To directly compare our results with those reported in Ref~1\cite{Exp}, we assign our structures the same labels as Ref~1. All the eleven conformers in Ref~1 have been identified in our simulations within an energy window of 0.2~eV from the global minimum. In addition, BOSS predicted new conformers, which we named N1, N2$\dots$. Some of them are shown in Figure \ref{fig:conformers}. The new conformers BOSS predicted generally have a higher energy.

\begin{figure*}[hbt!]
\includegraphics[width=17.5 cm]{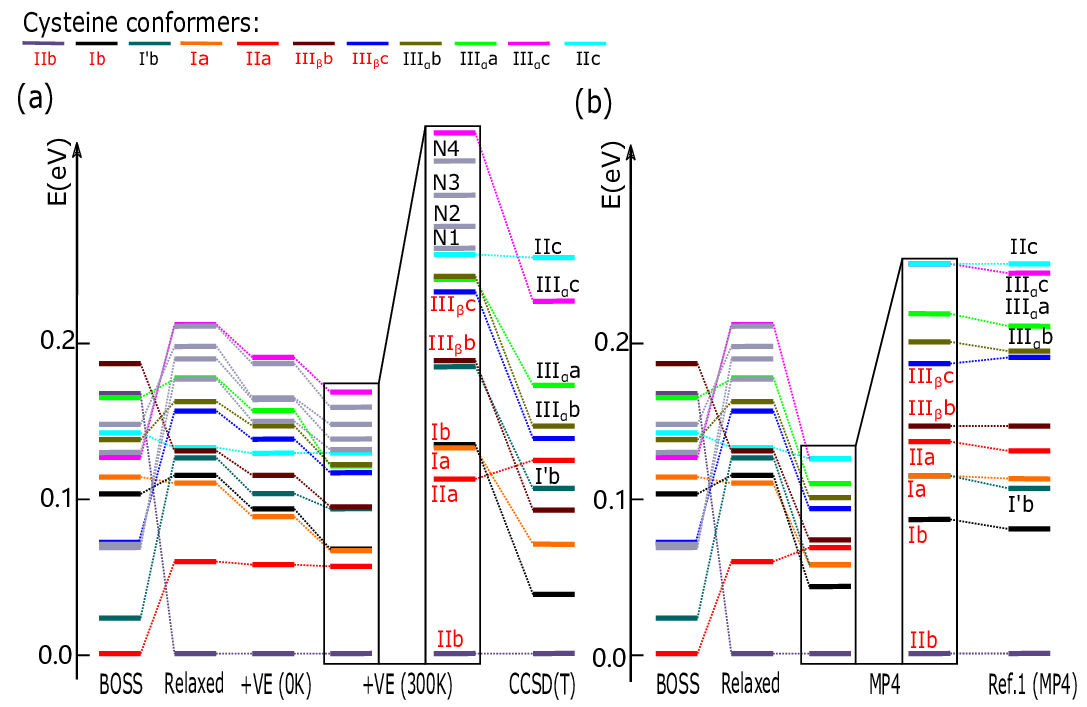}
\centering
\caption{ Relative stability for all steps of the PBE0+MBD-based search. 
(a) From left to right:  BOSS prediction, after structure optimization, after adding the vibrational energy at 0 K (+VE (0K)), and adding the vibration energy at 300 K (+VE (300K)). The two most right ones are +VE (300K) and the energy order of CCSD(T) result but enlarged 2 times. For each step, the energy of the most stable structure defines the zero of energy for that column.(b) From the left to right: BOSS prediction, after optimization, and MP4 energy calculations. The last two columns show an enlarged version of the MP4 results in comparison with the MP4 results of Ref~1\cite{Exp}.}
\label{fig:order}
\end{figure*}

The relative stability of the PBE0+MBD conformers is shown in Figure \ref{fig:order}a. Corresponding plots for the PBE+TS, PBE+MBD and the PBE0+TS functionals are presented in Figures S2-S4 of the Supporting Information. To illustrate the importance of different contributions to the energy hierarchy, Figure \ref{fig:order}a and Figure S2-S4 show not only the final energy order but also intermediate steps. 

The hierarchy figures show that once the conformers are extracted, geometry optimization plays a major role in refining their energy ranking. The largest energy changes and reordering happens in this step. In the PBE0+MBD simulation, the average energy change of the most stable 15 conformers during the geometry optimization is 0.095 eV, while the dihedral angles of the corresponding structures change on average by $\Delta d_1= 16.9^{\circ}$, $\Delta d_2= 20.9^{\circ}$, $\Delta d_3= 8.9^{\circ}$, $\Delta d_4= 26.1^{\circ}$ and $\Delta d_5= 11.9^{\circ}$.

The entropy corrections have a smaller effect on the conformer ordering. The zero-point energy contributions (+VE (0K) column) does not trigger any conformer reordering. It does, however, compress the energy spectrum as corrections for higher-energy conformers are larger than for the global minimum. The finite temperature corrections (+VE (300K) column) leads to a further compression of the energy spectrum. Now a couple of conformers above 0.1 eV switch orders as their vibrational entropy contributions differ.

The final column shows our most accurate conformer energy hierarchy, which now includes also the CCSD(T) corrections. We observe that the CCSD(T) corrections are sensitive to the conformer geometry. They generally shift conformers down in energy towards the global minimum conformer. This reduces the energy spacing between the conformers. Conformers IIa and IIc are an exception. They remain at the roughly the same relative energy to the global minimum, which is also of conformer type II. They subsequently trade places with other conformers in the hierarchy. 

The energy hierarchy of the low energy conformers from different methods are also listed in Table \ref{table:local-order}.

\subsection{Validation}

To validate our optimized conformer structures, we start with Ref~2\cite{Theo3-Cys}. The geometries reported in Ref~2 were obtained at the  MP2(FC)/aug-cc-pV(T$^+$d)Z level and we compare them against our PBE0+MBD geometries. To standardize the comparison, we use the same conformer naming convention as in Ref~1\cite{Exp}.

Among the top ten most stable structures,  Ref~2\cite{Theo3-Cys} reports eight structures that we and Ref~1 also found (see Tab.~\ref{table:local-energy}). These are IIb, IIa, Ib, I'b, Ia, III${_\beta}$b, III${_\beta}$c and III${_\alpha}$b.\footnote{Using  our  classification  standard,  we classified the  VI(n/a)  conformer  in  Ref~2\cite{Theo3-Cys} to be IIb.}The average differences of the dihedral angles between our and Ref~2's geometries are $\Delta d_1= 4.6^{\circ}$, $\Delta d_2= 1.4^{\circ}$, $\Delta d_3= 2.8^{\circ}$, $\Delta d_4= 0.7^{\circ}$ and $\Delta d_5= 3.0^{\circ}$. These small differences indicate that we indeed found the right conformers and that the PBE0+MBD and MP2 geometries agree closely.

Ref~1\cite{Exp}, unfortunately, does not provide atomic coordinates for the reported conformers. To validate our optimized conformer structures against those of Ref~1, we therefore performed MP4 single-point energy calculations with the same basis set 6-311++G(d,p) as in Ref~1\cite{Exp}, but for our PBE0+MBD geometries. The results are reported in Figure \ref{fig:order}b and Table \ref{table:local-order}. 

Figure \ref{fig:order}b and Table \ref{table:local-energy} show that the energies of the two MP4 calculations (MP4(b1) and MP4(b1)\cite{Exp}) agree within 4~meV for each conformer. This close match indicates that our conformer geometries agree very well with those of Ref~1, validating our BOSS-based conformer search procedure.

%\textcolor{red}{After finding the reliable conformer structures, we refine their energies by adding the free energy and coupled cluster corrections.} A few energy order changes also take place in these steps as shown in Figure \ref{fig:order}b. The relative vibrational energy change for structure IIa (red line) and IIc (cyan line) is small, whereas other conformers get shifted downwards by these energy corrections. IIa and IIc are also not affected much by the relative coupled cluster corrections, which again shift the other conformers down in energy closer to the ground state conformer.

Table \ref{table:local-order} shows the final energy ranking of the top ten most stable conformers in Ref~1\cite{Exp}, Ref~2\cite{Theo3-Cys} and our computational predictions. A more complete list of the low-energy conformers and their relative energy can be found in Table S1. 

In our simulations, PBE+TS, PBE+MBD, PBE0+TS and PBE0+MBD all found the correct global minimum structure  IIb. PBE+TS, PBE0+TS and PBE0+MBD predicted the six experimental identified conformers among the top seven most stable structures, while PBE+MBD locates the six conformers among the top eight most stable ones.

\begin{figure}[hbt!]
\includegraphics[width=7 cm]{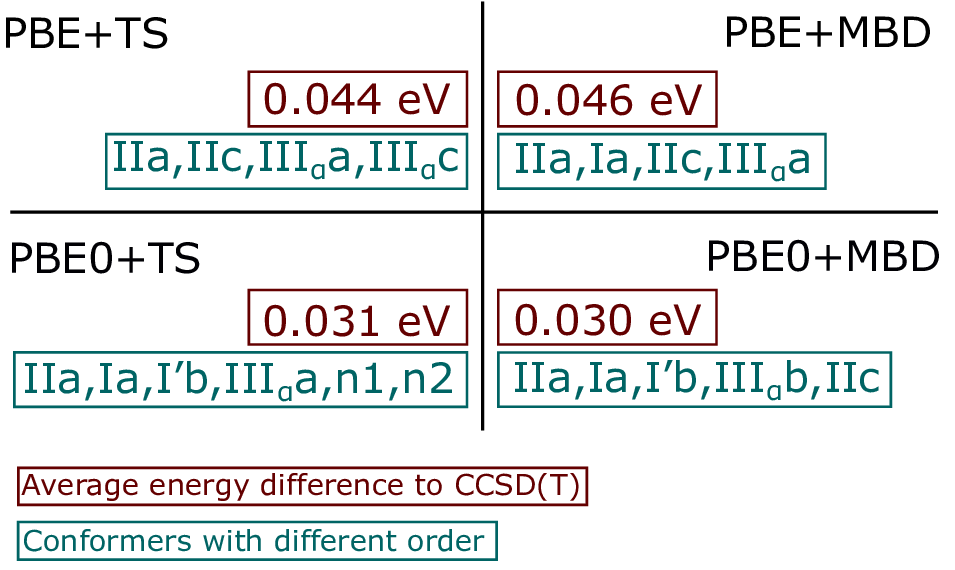}
\centering
\caption{\label{fig:dif} Summary of DFT results: each panel shows the average energy difference between the respective DFT functional and the CCSD(T) reference energies for the 10 lowest conformers. In addition, each panel lists the conformers that have a different order than in CCSD(T). }
\end{figure}

In Figure \ref{fig:dif}, we summarize the comparison across the four different exchange-correlation functionals we tested. Our reference are the CCSD(T) energies at the PBE0+MBD geometries. In Figure \ref{fig:dif} we list the conformers that have a different energy ordering in the DFT and CCSD(T). The energy differences between the cysteine conformers are extremely small. Therefore, it is no surprise that the DFT energy rankings differ from the CCSD(T) results. The accuracy of the different DFT functional are then evaluated by the energy differences comparing to CCSD(T), using the 10 lowest energy conformers in CCSD(T)). Comparing to CCSD(T), the average energy difference is 0.044 eV for PBE+TS, 0.046 eV for PBE+MBD, 0.031 eV for PBE0+TS and 0.030 eV for PBE0+MBD (Figure 8). PBE0 is on average 0.01~eV more accurate than PBE. The difference between the different van der Waals treatments (TS or MDB) is an order of magnitude smaller (1 or 2 meV on average).  The influence of the different vdW treatments is therefore negligible for a small molecule like cysteine. For cysteine, we can conclude that PBE+TS is sufficient for the conformer search.

Since BOSS is able to sample the configurational space very efficiently, we performed the whole conformer search at the PBE0+MBD level. For larger molecules, it might become more economical to perform an initial BOSS-based conformer search at the PBE+TS level and to post-relax only a certain number of low-energy conformers with PBE0+MBD.

Our CCSD(T) calculations produce a very similar energy ranking as the MP4 results in Ref~1 \cite{Exp}, as shown in Table \ref{table:local-order}. The only difference with Ref~1 is the placement of I$^{'}$b, III$_{\beta}$b.
If we use the same aug-cc-pvtz basis set, same geometries and same vibrational energy correction from our PBE0+MBD simulations in both CCSD(T) and MP4, we get the same energy order. Therefore, the differences are not caused by the choice of CCSD(T) or MP4. Since we have validated that we have found very similar structures as Ref~1\cite{Exp} (Figure \ref{fig:order}b), the difference may due to the facts that Ref~1\cite{Exp} did not include the entropy correction and used different basis set.

Ref~2\cite{Theo3-Cys} has found two new structures similar to IIa, not appearing in Ref~1\cite{Exp} and our results. Except for these two new structures, the only difference between our CCSD(T) and the CCSD(T) results in Ref~2\cite{Theo3-Cys} is the ordering of I$^{'}$b and III$_{\beta}$b. Again, the energy differences between the conformers in this range are extremely small, and ordering differences in our results and the reference can be ascribed to the slight difference of the conformer structures and computational settings. Ref~2\cite{Theo3-Cys} used a different vibrational correction method and included the Focal-Point analysis to extrapolate the energies to the complete basis set. 

Comparing our CCSD(T) results to experiment, we note that the CCSD(T) ordering of IIb, Ib, Ia as the three lowest energy conformers agrees with the experimental ordering derived from the relative abundance of the detected conformers. However the order of Ia and Ib is switched, the same as the computational ranking in Ref~1\cite{Exp} and Ref~2\cite{Theo3-Cys}.  For the next three conformers, the experiment finds IIa, III${_\beta}$b and III${_\beta}$c, however, with much lower overall abundance than the first three conformers. The coupled cluster order is different with III${_\beta}$b, I$^{'}$b, IIa and III${_\beta}$c. These differences can be ascribed to the low experimental abundance, which might make an unambiguous classification difficult, or to additional experimental factors that are not taken into account in our simulations.

\begin{table*}[hbt!]
\caption{The energy order of the ten most stable conformers of cysteine from our DFT, MP4 and CCSD(T) computations and Ref.~\citenum{Exp} (Our CCSD(T) and MP4 results are based on PBE0+MBD structures. b1: 6-311++G(d,p) basis set, b2: aug-cc-pvtz basis set, *: vibrational energy correction not included.)}

\begin{tabular}{lllllllllll}
\hline
               &  \multicolumn{10}{c}{Energy order}\\
\hline
 PBE+TS   & \textcolor{red}{IIb} & \textcolor{red}{IIa} & \textcolor{red}{Ib} &\textcolor{red}{Ia} &\textcolor{red}{$\mathrm{III_{\beta}b}$}  &I$^{'}$b & \textcolor{red}{$\mathrm{III_{\beta}c}$} & IIc & $\mathrm{III_{\alpha}a}$ & $\mathrm{III_{\alpha}b}$ \\
 PBE+MBD   & \textcolor{red}{IIb} &\textcolor{red}{IIa} & \textcolor{red}{Ia}& \textcolor{red}{Ib}& \textcolor{red}{$\mathrm{III_{\beta}b}$}& I$^{'}$b& IIc& \textcolor{red}{$\mathrm{III_{\beta}c}$}& $\mathrm{III_{\alpha}a}$& $\mathrm{III_{\alpha}b}$ \\
 PBE0+TS   &\textcolor{red}{IIb}  & \textcolor{red}{IIa}& \textcolor{red}{Ia}& \textcolor{red}{Ib}& I$^{'}$b& \textcolor{red}{$\mathrm{III_{\beta}b}$}& \textcolor{red}{$\mathrm{III_{\beta}c}$}&$\mathrm{III_{\alpha}a}$& N1&N2 \\
 PBE0+MBD  &\textcolor{red}{IIb}&\textcolor{red} {IIa} &\textcolor{red}{Ia} & \textcolor{red}{Ib}& I$^{'}$b& $\textcolor{red}{\mathrm{III_{\beta}b}}$& $\textcolor{red}{\mathrm{III_{\beta}c}}$& $\mathrm{III_{\alpha}a}$& $\mathrm{III_{\alpha}b}$&IIc \\
 MP4 (b1)$^*$  & \textcolor{red}{IIb} & \textcolor{red}{Ib} & \textcolor{red}{Ia} &I$^{'}$b&\textcolor{red}{IIa}& $\textcolor{red}{\mathrm{III_{\beta}b}}$&   $\textcolor{red}{\mathrm{III_{\beta}c}}$  & $\mathrm{III_{\alpha}b}$& $\mathrm{III_{\alpha}a}$& $\mathrm{III_{\alpha}c}$ \\
 MP4 (b1)$^*$ \cite{Exp}     & \textcolor{red}{IIb} & \textcolor{red}{Ib} &I$^{'}$b &\textcolor{red}{Ia} & \textcolor{red}{IIa} & $\textcolor{red}{\mathrm{III_{\beta}b}}$& $\textcolor{red}{\mathrm{III_{\beta}c}}$& $\mathrm{III_{\alpha}b}$& $\mathrm{III_{\alpha}a}$& $\mathrm{III_{\alpha}c}$\\
 MP4 (b2) & \textcolor{red}{IIb} & \textcolor{red}{Ib} & \textcolor{red}{Ia} & $\textcolor{red}{\mathrm{III_{\beta}b}}$ &I$^{'}$b&\textcolor{red}{IIa}&$\textcolor{red}{\mathrm{III_{\beta}c}}$&$\mathrm{III_{\alpha}b}$& $\mathrm{III_{\alpha}a}$&$\mathrm{III_{\alpha}c}$\\
  CCSD(T) (b2)         & \textcolor{red}{IIb} & \textcolor{red}{Ib} & \textcolor{red}{Ia} & $\textcolor{red}{\mathrm{III_{\beta}b}}$  &I$^{'}$b &  \textcolor{red}{IIa}& $\textcolor{red}{\mathrm{III_{\beta}c}}$  & $\mathrm{III_{\alpha}b}$& $\mathrm{III_{\alpha}a}$& $\mathrm{III_{\alpha}c}$  \\
 CCSD(T) \cite{Theo3-Cys}   & \textcolor{red}{IIb} & \textcolor{red}{Ib} &\textcolor{red}{Ia} & I$^{'}$b& \textcolor{red}{IIa} & $\textcolor{red}{\mathrm{III_{\beta}b}}$&n/a&n/a& $\textcolor{red}{\mathrm{III_{\beta}c}}$& $\mathrm{III_{\alpha}b}$\\ 
 
Exp \cite{Exp} & \textcolor{red}{IIb} & \textcolor{red}{Ia} & \textcolor{red}{Ib} &   \textcolor{red}{IIa} & $\textcolor{red}{\mathrm{III_{\beta}c}}$   & $\textcolor{red}{\mathrm{III_{\beta}b}}$ & \\
Abundance ratio\cite{Exp} & 10 & 10 & 8  & 3 & 3  &2\\
\hline
\end{tabular}
\label{table:local-order}
\end{table*}

%\begin{table*}[hbt!]
%\caption{Predicted low-energy conformers of cystine and relative energies with respect to the global minimum in eV (b1: 6-311++G(d,p) basis set, b2: aug-cc-pvtz basis set,*:  vibrational energy correction not included.}
%{Energy order of MP4: MP4(1)-(6-311++G(d,p)), MP4(2)-(aug-cc-pvtz), MP4(1V)-(MP4(1)+300 vibrational energy), MP4(2V)-(MP4(2)+300 vibrational energy)}
%\begin{tabular}{|c|c|c|c|c|c|c|c|c|c|c|c|}
%\hline
% Conformer & \textcolor{red}{IIb} & \textcolor{red}{Ib} &I$^{'}$b &\textcolor{red}{Ia} & \textcolor{red}{IIa} & $\textcolor{red}{\mathrm{III_{\beta}b}}$& $\textcolor{red}{\mathrm{III_{\beta}c}}$& $\mathrm{III_{\alpha}b}$& $\mathrm{III_{\alpha}a}$& $\mathrm{III_{\alpha}c}$&IIc\\
%\hline
%MP4 (b1)$^*$& 0.000 & 0.043 &0.057 & 0.057 & 0.068 & 0.073 & 0.093 & 0.100&0.109&0.125&0.125\\
%\hline
%MP4 (b1)$^*$\cite{Exp} & 0.000 & 0.040 & 0.053 & 0.056 & 0.065&0.073&0.095&0.097&0.105&0.122&0.125\\
%\hline
%  MP4 (b2) & 0.000 & 0.025 & 0.056 & 0.046 & 0.065 &0.054 &0.075 &0.079 &0.092&0.116&0.129\\
%\hline
 %CCSD(T) (b2) & 0.000 & 0.019 & 0.053 & 0.035 & 0.062 &0.046&0.069&0.073&0.086&0.113&0.127\\
 %\hline
%CCSD (T)\cite{Theo3-Cys} & 0.000 & 0.050 & 0.062 & 0.060 & 0.068&0.069&0.099&0.100&&&\\
%\hline
%
%\end{tabular}
%\label{table:local-energy}
%\end{table*}

\begin{table*}[hbt!]
\caption{Predicted low-energy conformers of cystine and relative energies with respect to the global minimum in eV (b1: 6-311++G(d,p) basis set, b2: aug-cc-pvtz basis set,*:  vibrational energy correction not included.}
\begin{tabular}{|c|c|c|c|c|c|c|c|c|c|c|c|}
\hline
 Conformer & \textcolor{red}{IIb} & \textcolor{red}{Ib}&\textcolor{red}{Ia}& $\textcolor{red}{\mathrm{III_{\beta}b}}$ &I$^{'}$b  & \textcolor{red}{IIa} & $\textcolor{red}{\mathrm{III_{\beta}c}}$& $\mathrm{III_{\alpha}b}$& $\mathrm{III_{\alpha}a}$& $\mathrm{III_{\alpha}c}$&IIc\\
 \hline
MP4 (b1)$^*$& 0.000 & 0.043 &0.057& 0.073 & 0.057 & 0.068  & 0.093 & 0.100&0.109&0.125&0.125\\
\hline
MP4 (b1)$^*$\cite{Exp} & 0.000 & 0.040 & 0.056&0.073& 0.053  & 0.065&0.095&0.097&0.105&0.122&0.125\\
\hline
  MP4 (b2) & 0.000 & 0.025& 0.046  &0.054&0.056  & 0.065  &0.075 &0.079 &0.092&0.116&0.129\\
\hline
 CCSD(T) (b2) & 0.000 & 0.019 & 0.035&0.046& 0.053  & 0.062 &0.069&0.073&0.086&0.113&0.127\\
 \hline
CCSD (T)\cite{Theo3-Cys} & 0.000 & 0.050& 0.060&0.069& 0.062 & 0.068  &0.099&0.100&&&\\
\hline

\end{tabular}
\label{table:local-energy}
\end{table*}

\subsection{Computational efficiency}
We close with a note on the efficiency of our new conformer search procedure without explicitly performing other search methods in this work. BOSS predicts a physically meaningful 5-D PES with only 1000 single-point DFT calculations. We can put this number of single-point calculations into perspective, by considering that FHI-aims requires on average 30 geometry optimization steps to relax the structure of cysteine. The computational cost of 1000 single-point DFT calculations  is therefore equivalent to approx. 30 DFT geometry optimizations.

From the PES, we extract all relevant low-energy conformers with the build-in BOSS minima tool at no further computational expense. In this work, we consider approx.~30 local minima, each of which is geometry optimized with DFT. This amounts to 30  geometry optimizations, which is equivalent to  approx.~900 DFT single-point calculations.

Our total computational expense per DFT functional for a complete conformer search of cysteine is therefore 1900 DFT single-point calculations or equivalently 60 geometry optimizations. This is a very small computational budget, compared to systematic \cite{Theo3-Cys} or stochastic \cite{ML-M2} conformer search methods that need to relax thousands of structures. Supady \textit{et al.} provide detailed numbers for a genetic algorithm (GA) based conformer search of dipeptides \cite{ML-M2}. Their search encompasses between 4 and 6 degress of freedom and is therefore similar to ours, as is the size of the molecules. The GA search requires between 20,000 and 60,000 single-point DFT calculations (referred to as force evaluations in Ref.~\citenum{ML-M2}) depending on the size of the search space and the density of conformers in the energy hierarchy. Our BOSS-based procedure is a factor of 10 more efficient. A similar speed-up was recently observed in a Gaussian-process-based structure search of oxidized graphene on the Ir(111)\cite{Ir111}.

\section{Conclusions}
In summary, we propose a new conformer search procedure that combines the Bayesian optimization active learning with quantum chemistry methods.
BOSS find all the relevant conformers in one run, with all the experimental ones already list within the lowest energy conformers. Then we refine the low-energy conformers by DFT structure relaxation, vibrational energy, and coupled cluster correction. We conclude that the DFT structure relaxation plays a major role in the refinement of the energy order. We also find that PBE0 gives slightly better results than PBE, but the difference between the TS and MBD van der Waals interactions are tiny for our system.

Unlike traditional conformer search methods, our approach is computationally tractable while retaining the accuracy of the chosen quantum chemical method throughout. The method is straightforward to apply to conformer search of other molecules. Besides the local minima structures and energies, our procedure also predicts the whole energy landscapes. In future work, the converged energy landscapes can be additional data-mined for energy barriers between pairs of conformers.

%%%%%%%%%%%%%%%%%%%%%%%%%%%%%%%%%%%%%%%%%%%%%%%%%%%%%%%%%%%%%%%%%%%%%
%% The "Acknowledgement" section can be given in all manuscript
%% classes.  This should be given within the "acknowledgement"
%% environment, which will make the correct section or running title.
%%%%%%%%%%%%%%%%%%%%%%%%%%%%%%%%%%%%%%%%%%%%%%%%%%%%%%%%%%%%%%%%%%%%%
\begin{acknowledgement}
This work was supported by the Academy of Finland (Project numbers  308647, 314298 and 316601) and through their Flagship programme: Finnish Center for Artificial Intelligence FCAI. We thank CSC, the Finnish IT Center for Science  and Aalto Science IT for computational resources. This work is supported by COST (European Cooperation in Science and Technology) Action 18234. Lincan Fang thanks 
Guoxu Zhang, Marc Dvorak, Jingrui Li and Annika Stuke for help with FHI-Aims. He also acknowledges financial support from Chinese Scholarship Council (Grant No. [2017]3109). 
\end{acknowledgement}

%%%%%%%%%%%%%%%%%%%%%%%%%%%%%%%%%%%%%%%%%%%%%%%%%%%%%%%%%%%%%%%%%%%%%
%% The same is true for Supporting Information, which should use the
%% suppinfo environment.
%%%%%%%%%%%%%%%%%%%%%%%%%%%%%%%%%%%%%%%%%%%%%%%%%%%%%%%%%%%%%%%%%%%%%
\begin{suppinfo}

%This will usually read something like: ``Experimental procedures and
%characterization data for all new compounds. 
%The class will
%automatically add a sentence pointing to the information on-line:

\end{suppinfo}

%%%%%%%%%%%%%%%%%%%%%%%%%%%%%%%%%%%%%%%%%%%%%%%%%%%%%%%%%%%%%%%%%%%%%
%% The appropriate \bibliography command should be placed here.
%% Notice that the class file automatically sets \bibliographystyle
%% and also names the section correctly.
%%%%%%%%%%%%%%%%%%%%%%%%%%%%%%%%%%%%%%%%%%%%%%%%%%%%%%%%%%%%%%%%%%%%%
\bibliography{achemso-demo}

\end{document}

\section{Table of Contents}
S1: The energy order of the Cysteine's conformers\\
S2: Eleven geometries from PBE0+MBD search. The label is same with the Ref.1 and coordinates are in angstrom.
\section{Figure of Contents}
S1: The 2-D (d1, d2) projected PES maps predicted by BOSS in 5-D case of cysteine.\\
S2: Relative stability for all steps of the PBE+TS-based search.\\
S3: Relative stability for all steps of the PBE+MBD-based search\\
S4: Relative stability for all steps of the PBE0+TS-based search\\
S5: The comparison between results of DFT and CCSD(T) simulation.\\

\begin{table}[htb]
\caption{The energy order of the low energy conformers of cysteine }
\begin{tabular}{llllllllllll}
\hline
\multicolumn{11}{c}{Different calculations}\\
\hline
\multicolumn{2}{c}{Ref.1 }& 
\multicolumn{2}{c}{CCSD(T)}&
\multicolumn{2}{c}{PBE+TS}& \multicolumn{2}{c}{PBE+MBD} & \multicolumn{2}{c}{PBE0+TS} & \multicolumn{2}{c}{PBE0+MBD}\\
\hline
Type&E (eV)&Type&E (eV)&
Type&E (eV)&Type&E (eV)&
Type&E (eV)&Type&E (eV)\\
\hline
\textcolor{red}{IIb}& 0.0&\textcolor{red}{IIb}& 0.0&
\textcolor{red}{IIb}& 0.0 &\textcolor{red}{IIb}& 0.0&
\textcolor{red}{IIb}& 0.0&\textcolor{red}{IIb}&0.0\\

\textcolor{red}{Ib}& 0.040&\textcolor{red}{Ib}& 0.019&
\textcolor{red}{IIa}& 0.051&\textcolor{red}{IIa}& 0.051&
\textcolor{red}{IIa}& 0.056&\textcolor{red}{IIa}& 0.056\\

I$'$b &0.053&\textcolor{red}{Ia}&0.035&
\textcolor{red}{Ib}&0.086&\textcolor{red}{Ia}&0.089&
\textcolor{red}{Ia}&0.065&\textcolor{red}{Ia}&0.066\\

\textcolor{red}{Ia}&0.056&\textcolor{red}{$\mathrm{III_{\beta}b}$}& 0.046&
\textcolor{red}{Ia}&0.087&\textcolor{red}{Ib}&0.089&
\textcolor{red}{Ib}&0.067&\textcolor{red}{Ib}&0.067\\

\textcolor{red}{IIa}&0.065&I$'$b&0.053&
\textcolor{red}{$\mathrm{III_{\beta}b}$}&0.115&\textcolor{red}{$\mathrm{III_{\beta}b}$}&0.117&
I$'$b&0.091&I$'$b&0.092\\

\textcolor{red}{$\mathrm{III_{\beta}b}$}&0.073&\textcolor{red}{IIa}&0.062&
I$'$b& 0.118&I$'$b& 0.121&
\textcolor{red}{$\mathrm{III_{\beta}b}$}&0.094&\textcolor{red}{$\mathrm{III_{\beta}b}$}&0.094\\

\textcolor{red}{$\mathrm{III_{\beta}c}$}&0.095&\textcolor{red}{$\mathrm{III_{\beta}c}$}& 0.069&
\textcolor{red}{$\mathrm{III_{\beta}c}$}&0.126&IIc&0.124&
\textcolor{red}{$\mathrm{III_{\beta}c}$}&0.111&\textcolor{red}{$\mathrm{III_{\beta}c}$}&0.116\\

 $\mathrm{III_{\alpha}b}$ &0.097&$\mathrm{III_{\alpha}b}$&0.073&
IIc&0.126&\textcolor{red}{$\mathrm{III_{\beta}c}$}&0.133&
$\mathrm{III_{\alpha}a}$&0.120&$\mathrm{III_{\alpha}a}$&0.120\\

$\mathrm{III_{\alpha}a}$&0.105&$\mathrm{III_{\alpha}a}$&0.086&
$\mathrm{III_{\alpha}a}$&0.135&$\mathrm{III_{\alpha}a}$&0.133&
N1&0.132&$\mathrm{III_{\alpha}b}$&0.121\\

$\mathrm{III_{\alpha}c}$&0.122&$\mathrm{III_{\alpha}c}$&0.113&
$\mathrm{III_{\alpha}b}$&0.140&$\mathrm{III_{\alpha}b}$&0.146&
N2&0.143&IIc&0.128\\

IIc&0.125&IIc&0.127&
N1&0.157&N1&0.156&
N4&0.150&N1&0.130\\

&&&&
N4&0.165&
N2&0.162&
IIc&0.150&
N2&0.137\\

&&&&
N2&0.167&
N3&0.170&
$\mathrm{III_{\alpha}c}$&0.158&
N3&0.147\\

&&&&
$\mathrm{III_{\alpha}c}$&0.173&
N4&0.173&
N5&0.183&
N4&0.158\\

&&&&
N5&0.203&
$\mathrm{III_{\alpha}c}$&0.184&
N6&0.188&
$\mathrm{III_{\alpha}c}$&0.167\\

\hline
\end{tabular}
\begin{tabbing}
$^\dag$ The experimental detected conformers are marked by red. 
\end{tabbing}
\end{table}

\begin{figure}[h]
\includegraphics[width=16.5cm]{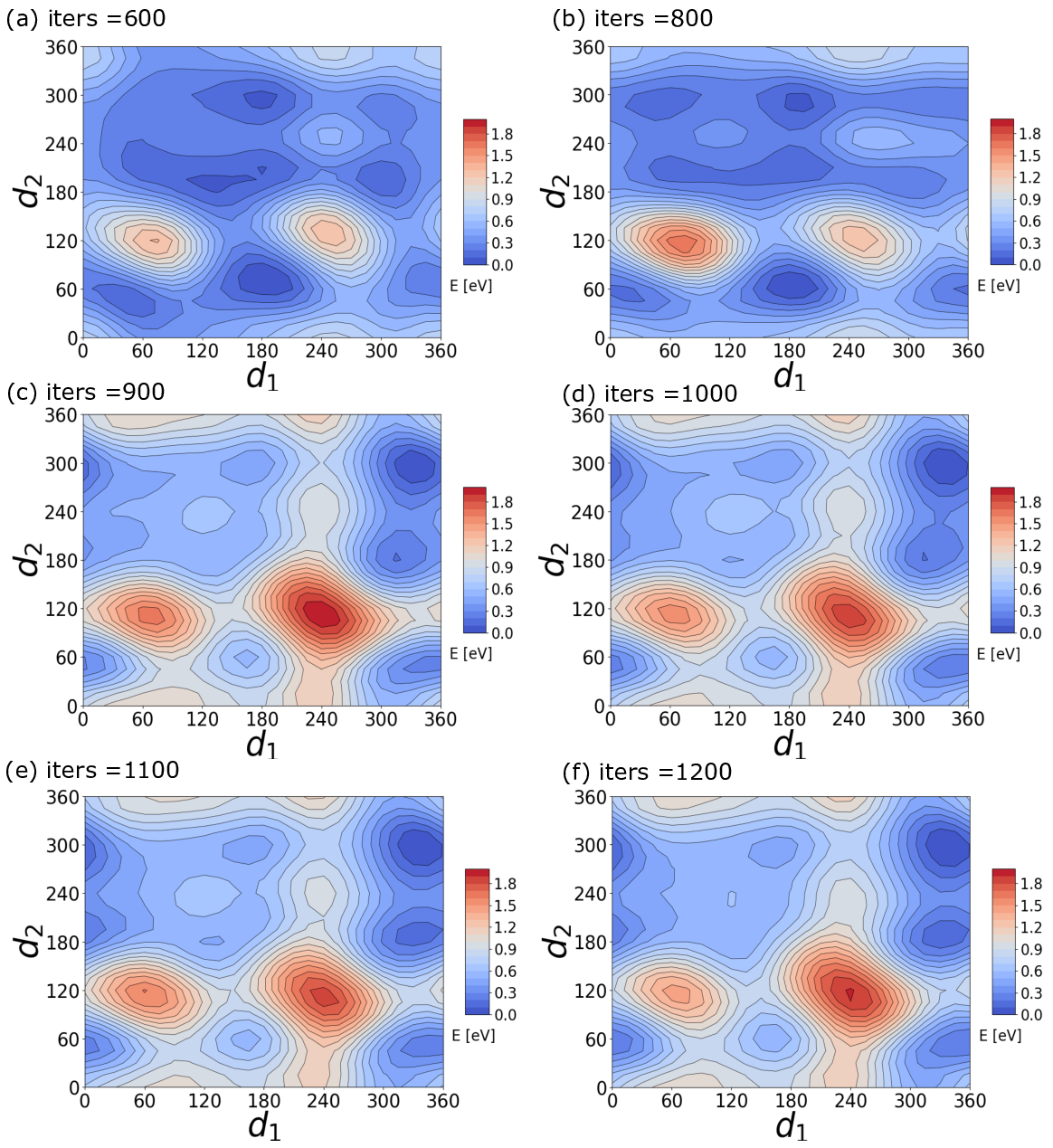}
\centering
\caption{The 2-D (d1, d2) projected PES maps predicted by BOSS in 5-D case of cysteine from (a) 600 iterations, (b) 800 iterations, (c) 900 iterations, (d) 1000 iterations, (e) 1100 iterations, and (f) 1200 iterations. The PES maps become very similar after 900 iterations indicating the BOSS prediction of the whole PES has converged.}
\end{figure}

\begin{figure}[h]
\includegraphics[width=16cm]{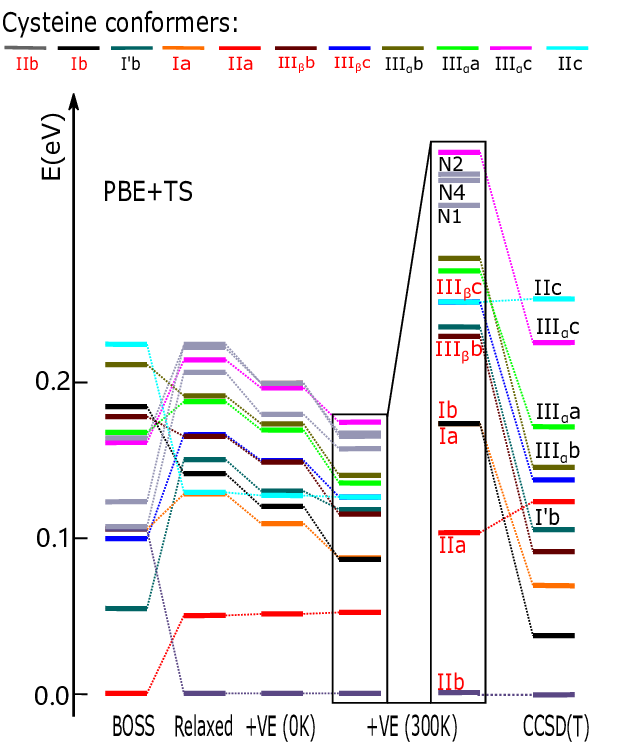}
\centering
\caption{Relative stability for all steps of the PBE+TS-based search. From left to right: 
BOSS prediction, after structure optimization, after adding vibration energy at 0 K (+VE (0K)) and after adding vibration energy at 300 K (+VE (300K)). The two most right ones are +VE (300K) and the energy order of CCSD(T) result but enlarged 2 times. For each step, the energy of the most stable structure defines the zero of the energy.}
\end{figure}

\begin{figure}[h]
\includegraphics[width=16cm]{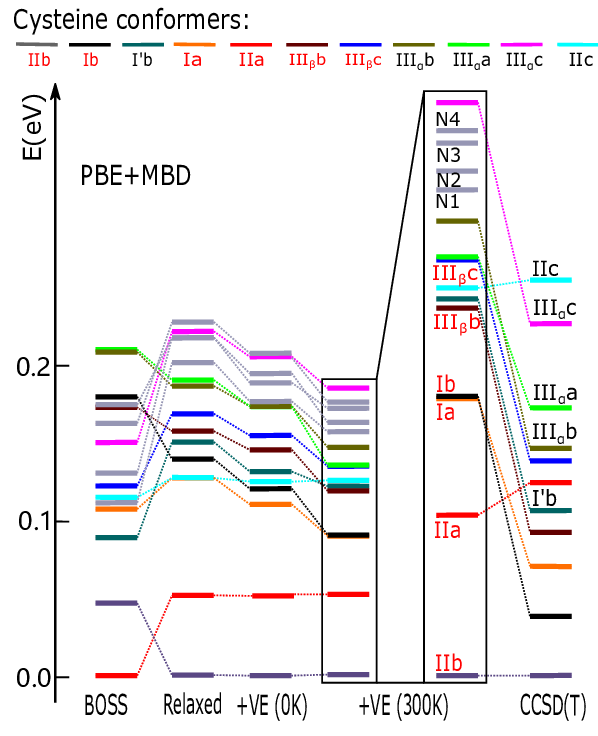}
\centering
\caption{Relative stability for all steps of the PBE+MBD-based search. From left to right: 
BOSS prediction, after structure optimization, after adding vibration energy at 0 K (+VE (0K)) and after adding vibration energy at 300 K (+VE (300K)). The two most right ones are +VE (300K) and the energy order of CCSD(T) result but enlarged 2 times. For each step, the energy of the most stable structure defines the zero of the energy.}
\end{figure}

\begin{figure}[h]
\includegraphics[width=16cm]{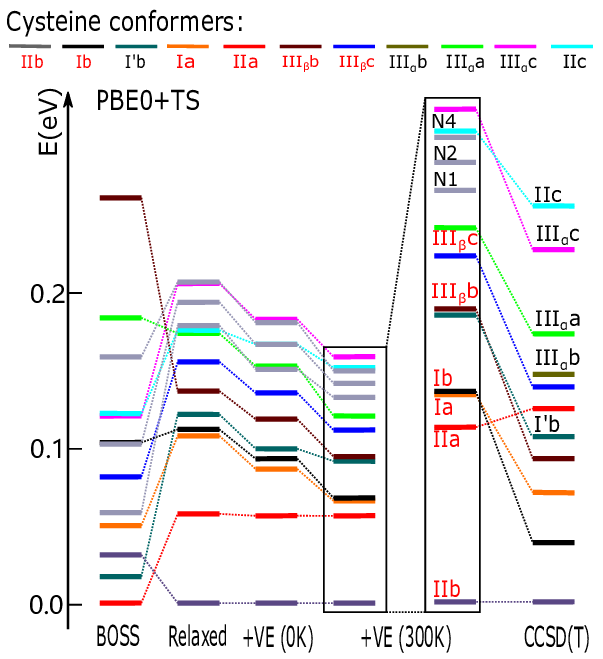}
\centering
\caption{Relative stability for all steps of the PBE0+TS-based search. From left to right: 
BOSS prediction, after structure optimization, after adding vibration energy at 0 K (+VE (0K)) and after adding vibration energy at 300 K (+VE (300K)). The two most right ones are +VE (300K) and the energy order of CCSD(T) result but enlarged 2 times. For each step, the energy of the most stable structure defines the zero of the energy.}
\end{figure}

\clearpage

\begin{longtable}{llllllllllll}
    \caption{Eleven geometries from PBE0+MBD search. The label is same with the Ref.1 and coordinates are in angstrom.}
\hline
%\begin{tabular}{llllllllllll}
IIb\\
N&	 -0.00658720  &   -0.09596219 &   -0.06312148 \\
C&	  1.43779486   &   0.04648462  &   -0.00345070 \\
C&	  1.88308090  &    0.11146695  &    1.46464050 \\
O&	  0.96162966  &   -0.32020762   &   2.31246330 \\
O	&  2.96838410   &   0.49214876 &     1.80052240 \\
C	&  2.02312607  &    1.21283124 &    -0.78929042 \\
S&	  1.29765951   &   2.81123546  &   -0.34683742 \\
H&	 -0.30485908 &    -0.62585161 &    -0.86787671 \\
H&	 -0.43771199   &   0.82215649   &  -0.10372445 \\
H	&  1.88769564  &   -0.87570173 &    -0.38749435 \\
H	&  0.17770287  &   -0.50606947 &     1.75138113 \\
H	&  3.10089758   &   1.25282253 &    -0.64854396 \\
H	&  1.81705375  &    1.08540061 &    -1.85250000 \\
H&	  1.86850329    &  2.89720596  &    0.86528214 \\
\hline
Ib\\
N&	  0.00895578 &     0.02010036&     -0.03985242 \\
C&	  1.44319699  &    0.03714770&      0.04407407 \\
C&	  1.97399466   &   0.01199118 &     1.47212974 \\
O&	  3.31351696   &   0.12542543 &     1.51745098 \\
O&	  1.30242155   &  -0.12807614  &    2.45367281 \\
C&	  2.01953431   &   1.20323458   &  -0.75269042 \\
S&	  1.51151366   &   2.83706922    & -0.15784672 \\
H&	 -0.34810289   &   0.91885751&      0.26460912 \\
H&	 -0.37008772   &  -0.66781824 &     0.59675279 \\
H&	  1.83856960   &  -0.87638936  &   -0.41999069 \\
H&	  3.57075480   &   0.05415655   &   2.44452537 \\
H&	  3.10434784   &   1.15189729&     -0.79238206 \\
H&	  1.63205139   &   1.14210449 &    -1.76913405 \\
H&	  2.26316306   &   2.84392944  &    0.95385147 \\
\hline
Ia\\
N&	 -0.01048138&     -0.00721551&     -0.03216058 \\
C&	  1.42809125 &     0.06423070 &     0.02782292 \\
C&	  1.95635499  &    0.13295611  &    1.45225499 \\
O&	  3.26680527   &  -0.14361199   &   1.51686952 \\
O&	  1.30099437    &  0.42736913    &  2.41257465 \\
C&	  1.93806475&      1.29707036&     -0.72592153 \\
S&	  1.56570469 &     1.23119421 &    -2.48932682 \\
H&	 -0.41058446  &    0.59347500  &    0.67851365 \\
H&	 -0.33393833   &  -0.94335709   &   0.16295400 \\
H&	  1.85690503&     -0.81787050    & -0.45109958 \\
H&	  3.53512472 &    -0.04467068&      2.43842205 \\
H&	  1.50690117  &    2.19821835 &    -0.28503421 \\
H&	  3.02222136   &   1.36634610  &   -0.64519087 \\
H&	  0.27066657   &   0.92109580   &  -2.31305821 \\
\hline
$\mathrm{III_{\beta}b}$\\
N&	  0.00642381&     -0.05563911&     -0.11858266 \\
C&	  1.43947635 &     0.03450782 &     0.02437781 \\
C&	  1.97397884  &    0.05156366  &    1.45277379 \\
O&	  1.22170259   &  -0.68502214   &   2.28392043 \\
O&	  2.97195819    &  0.61330955    &  1.81348612 \\
C&	  1.98635195&      1.21943618     &-0.75573286 \\
S&	  1.34104380 &     2.82285183&     -0.21097749 \\
H&	 -0.41763711  &    0.80797438 &     0.19879112 \\
H&	 -0.36883468   &  -0.80804099  &    0.44009366 \\
H&	  1.88186325    & -0.86939132   &  -0.41731103 \\
H&	  1.64766344 &    -0.65381306    &  3.14880068 \\
H&	  3.07211341  &    1.24172297&     -0.71379050 \\
H&	  1.67532854   &   1.11973934 &    -1.79450392 \\
H&	  2.07132762    &  2.89704089  &    0.91175486 \\
\hline
I$^'$b\\
N&	  0.00320451&      0.02296452&     -0.02559052 \\
C&	  1.44683661 &     0.02961707 &    -0.01015214 \\
C&	  1.93092652  &   -0.00479547  &    1.42492759 \\
O&	  3.25640217   &   0.14792497   &   1.51323559 \\
O&	  1.22526609    & -0.19054875    &  2.37629019 \\
C&	  2.00889031 &     1.23401077&     -0.76634279 \\
S&	  1.53204184 &     2.80941763 &    -0.02391225 \\
H&	 -0.35153276  &   -0.28584536  &    0.86972559 \\
H&	 -0.36402238   &  -0.57090559   &  -0.75181911 \\
H&	  1.87893066    & -0.86914996    & -0.47619823 \\
H&	  3.48221606 &     0.11776926&      2.45015353 \\
H&	  3.09738522  &    1.20961385 &    -0.76582134 \\
H&	  1.66617084   &   1.19119204  &   -1.80039574 \\
H&	  0.22502430    &  2.50247504   &   0.03507963 \\
\hline
IIa\\
N&	 -0.02531896 &    -0.02038687&     -0.01838661 \\
C&	  1.42642687  &   -0.02615619 &     0.00825131 \\
C&	  1.92396416   &   0.07115776  &    1.45496568 \\
O&	  0.98858825    &  0.46893255   &   2.30925311 \\
O&	  3.05750934     &-0.15293486    &  1.76550325 \\
C&	  1.99117309&      1.18377033     &-0.73222599 \\
S&	  1.50153517 &     1.28744805&     -2.46924258 \\
H&	 -0.39967726  &   -0.95596300 &     0.05458059 \\
H&	 -0.35433288   &   0.36917824  &   -0.89432994 \\
H&	  1.87164836    & -0.93102643   &  -0.41672290 \\
H&	  0.17344030&      0.54703572    &  1.76870461 \\
H&	  1.61063323 &     2.10394856     &-0.28307856 \\
H&	  3.07463297  &    1.19072170&     -0.63960056 \\
H&	  2.12374737   &   0.17041444 &    -2.87763142 \\
\hline
$\mathrm{III_{\beta}c}$\\
N&	 -0.01005020&      0.03114401 &     0.01584135 \\
C&	  1.44568303 &    -0.00725444  &   -0.00102187 \\
C&	  1.98586250  &   -0.02457748   &   1.41832711 \\
O&	  2.01820205   &  -1.26656774    &  1.92371615 \\
O&	  2.28076837    &  0.94675444     & 2.06113784 \\
C&	  1.97462402&      1.18757503&     -0.76963281 \\
S&	  3.76648463 &     1.18079089 &    -0.99986473 \\
H&	 -0.34216014  &    0.81678150  &    0.56057017 \\
H&	 -0.39333298   &  -0.81202022   &   0.42017790 \\
H&	  1.75636581    & -0.92664536    & -0.49843227 \\
H&	  2.30213925&     -1.19241344&      2.84304018 \\
H&	  1.53763344 &     1.17339578 &    -1.76719586 \\
H&	  1.67653464  &    2.11599446  &   -0.28292915 \\
H&	  4.08005558   &   1.52599258   &   0.25747598 \\
\hline
$\mathrm{III_{\alpha}b}$\\
N&	  0.02632765&     -0.12698912&     -0.17318104 \\
C&	  1.45397687 &    -0.00794831 &    -0.01713689 \\
C&	  2.00554110  &   -0.05762961  &    1.40616329 \\
O&	  1.08926468   &  -0.41959002   &   2.31462263 \\
O&	  3.14927957    &  0.16653930    &  1.69041361 \\
C&	  1.98023849     & 1.24161656     &-0.71262203 \\
S&	  1.27572264&      2.78399356&     -0.07493115 \\
H&	 -0.43718703 &     0.59829764 &     0.35913208 \\
H&	 -0.30319170  &   -1.00978014  &    0.18968175 \\
H&	  1.92355858   &  -0.86266219   &  -0.51916316 \\
H&	  1.54035433    & -0.43742187    &  3.16738648 \\
H&	  3.05336330 &     1.32349290     &-0.55168454 \\
H&	  1.78667496  &    1.17441433&     -1.78071957 \\
H&	  0.09764656   &   2.69380695 &    -0.70974147 \\
\hline
$\mathrm{III_{\alpha}a}$\\
N&	 -0.03599748&      0.12647711&     -0.02817196 \\
C&	  1.40495123 &     0.05739079 &     0.03530189 \\
C&	  2.01841644  &    0.11966675  &    1.42624040 \\
O&	  1.28585202   &   0.85151587   &   2.28382552 \\
O&	  3.06062255    & -0.38722604    &  1.73222846 \\
C&	  2.02982668     & 1.18795919     &-0.78641218 \\
S&	  1.64803014 &     1.06096434&     -2.54517630 \\
H&	 -0.38556076  &    0.89498463 &     0.52806931 \\
H&	 -0.46128403   &  -0.72021009  &    0.31947384 \\
H&	  1.73384420    & -0.88908207   &  -0.39483972 \\
H&	  1.77012022     & 0.87193239    &  3.11787987 \\
H&	  1.69904095&      2.15341844     &-0.39776459 \\
H&	  3.11578741 &     1.14555101&     -0.70586827 \\
H&	  0.33276043  &    0.86314767 &    -2.35868628 \\
\hline
$\mathrm{III_{\alpha}c}$\\
N&	 -0.01580408&     -0.01350031&     -0.03083343 \\
C&	  1.43881177 &     0.00312142 &     0.00621186 \\
C&	  1.93761811  &   -0.04221132  &    1.44010213 \\
O&	  1.89812519   &   1.16724462   &   2.03535322 \\
O&	  2.25688823    & -1.03737570    &  2.02602902 \\
C&	  1.94282100     & 1.20577647     &-0.77251161 \\
S&	  3.74182902 &     1.30643341&     -0.90682139 \\
H&	 -0.40077008  &    0.76552677 &     0.48735875 \\
H&	 -0.37919837   &  -0.86603803  &    0.37255840 \\
H&	  1.80423481    & -0.90982816   &  -0.46353270 \\
H&	  2.15309964     & 1.03284685    &  2.95589189 \\
H&	  1.57027558 &     1.12631794     &-1.79339965 \\
H&	  1.55468383  &    2.13230490    & -0.35047912 \\
H&	  3.97468534   &   1.70637114     & 0.35174263 \\
\hline
IIc \\
N&	  0.01452693&     -0.09881227&      0.00139358 \\
C&	  1.46520528 &     0.06021291 &    -0.01120138 \\
C&	  1.97696731  &   -0.02843766  &    1.43287703 \\
O&	  1.14405212   &  -0.66547802   &   2.24646307 \\
O&	  3.03159941    &  0.40901730    &  1.79623689 \\
C&	  1.96557966     & 1.31411819     &-0.72138624 \\
S&	  3.69198404&      1.25610564&     -1.25142617 \\
H&	 -0.32631260 &    -0.47805241 &    -0.86943918 \\
H&	 -0.45175756  &    0.78594610  &    0.15684239 \\
H&	  1.88771190   &  -0.81252459   &  -0.51854525 \\
H&	  0.34324030    & -0.83423064    &  1.70871969 \\
H&	  1.39635227     & 1.44051970     &-1.64346142 \\
H&	  1.79642123      &2.19634399    & -0.10129336 \\
H&	  4.19838969     & 1.11383175     &-0.01687964\\
\hline
N1\\
N&	  0.04432511&     -0.01215031&     -0.04793693 \\
C&	  1.48098734 &     0.03337623 &    -0.00293661 \\
C&	  1.92603949  &    0.00283641  &    1.44466062 \\
O&	  3.22171612   &   0.32135797   &   1.58710806 \\
O&	  1.23062423    & -0.31836508    &  2.36585149 \\
C&	  2.00407551     & 1.26090329&     -0.74274409 \\
S&	  1.57829324      &1.25543375 &    -2.50253846 \\
H&	 -0.32050903&     -0.65859832  &    0.63615792 \\
H&	 -0.27861818 &    -0.26019085   &  -0.97083413 \\
H&	  1.97477257  &   -0.84992966    & -0.44707081 \\
H&	  3.42954524   &   0.22812140&      2.52434611 \\
H&	  1.52304609    &  2.15171346 &    -0.33760185 \\
H&	  3.08002291     & 1.36453419  &   -0.63014454 \\
H&	  2.46673935     & 0.32078752   &  -2.87425677\\
\\
\hline
N2\\
N&	  0.00675012&     -0.02453557&     -0.04581232 \\
C&	  1.45248063 &     0.01773015 &    -0.00523981 \\
C&	  1.93905461  &   -0.00054370  &    1.43081588 \\
O&	  1.28794725   &  -0.92090743   &   2.15858190 \\
O&	  2.82225850    &  0.67484353    &  1.87757493 \\
C&	  1.97137943 &     1.25752696     &-0.71228274 \\
S&	  1.55178972  &    1.30086442&     -2.47158020 \\
H&	 -0.34782849   &  -0.85806242 &     0.39960467 \\
H&	 -0.29744752    & -0.01913875  &   -1.01093990 \\
H&	  1.93149212     &-0.86324904   &  -0.47005599 \\
H&	  1.65794356 &    -0.88968582    &  3.04881113 \\
H&	  1.49901351  &    2.14132438&     -0.28279549 \\
H&	  3.04566260   &   1.35548054 &    -0.58063349 \\
H&	  2.37569396    &  0.31406276  &   -2.85727857\\
\\
\hline
N3\\
N&	 -0.01327853&      0.01586804&      0.03071639 \\
C&	  1.42423277 &     0.06162765 &     0.00581150 \\
C&	  1.93792051  &    0.02520575  &    1.42821935 \\
O&	  3.26654733   &  -0.14428015   &   1.46742137 \\
O&	  1.25913516    &  0.13697244    &  2.40801648 \\
C&	  2.03669734     & 1.24928186&     -0.74850020 \\
S&	  1.77243666&      2.86496295 &     0.01661526 \\
H&	 -0.42853003 &     0.44491804  &   -0.78160194 \\
H&	 -0.37341213  &    0.43864379   &   0.87494238 \\
H&	  1.81553119   &  -0.83882817    & -0.48574464 \\
H&	  3.52563292    & -0.12321587     & 2.39598185 \\
H&	  3.11923042 &     1.13418891&     -0.80290761 \\
H&	  1.65366010  &    1.25255911 &    -1.77011999 \\
H&	  0.43242630   &   2.80241563  &    0.05327978 \\
\\
\hline
N4\\
N&	 -0.00495413&      0.03423137&      0.04305660 \\
C&	  1.44870970 &    -0.00294695 &    -0.03272255 \\
C&	  1.99456613  &   -0.03023909  &    1.38096382 \\
O&	  1.49016637   &  -1.06489884   &   2.07239817 \\
O&	  2.77818457    &  0.74254847    &  1.85444452 \\
C&	  1.95216166 &     1.21181499     &-0.79375309 \\
S&	  3.68840005  &    1.13012094&     -1.28345716 \\
H&	 -0.35322109   &  -0.72443765 &     0.61238963 \\
H&	 -0.40323022    & -0.06684625  &   -0.88015809 \\
H&	  1.84321258 &    -0.90686388   &  -0.52172101 \\
H&	  1.86559518  &   -1.02281410    &  2.95989512 \\
H&	  1.39872342   &   1.28760261&     -1.73106950 \\
H&	  1.76172966    &  2.11911555 &    -0.22365007 \\
H&	  4.19182614     & 1.27019281  &   -0.04897638 \\
\\
\hline
N5\\
N&	  0.04701095&     -0.07015259&      0.08345191 \\
C&	  1.49256653 &     0.02095577 &    -0.03337658 \\
C&	  2.02695658  &   -0.00125043  &    1.38707254 \\
O&	  2.42267938   &  -1.21483705   &   1.77283874 \\
O&	  2.05984442    &  0.96286338    &  2.10336318 \\
C&	  1.98135872     & 1.26802609&     -0.76050431 \\
S&	  3.77833140&      1.36500609 &    -0.95966694 \\
H&	 -0.38189789 &    -0.19161037  &   -0.82277568 \\
H&	 -0.33041788  &    0.76973181   &   0.50269478 \\
H&	  1.85430813   &  -0.87073761    & -0.54571509 \\
H&	  2.69438747 &    -1.14580829&      2.69650098 \\
H&	  1.57288004  &    1.27040717 &    -1.77203001 \\
H&	  1.62437994   &   2.16173045  &   -0.24898223 \\
H&	  4.04956222    &  1.69551558   &   0.31157870 \\
\\
\hline
\end{longtable}